\documentclass[
 reprint,
superscriptaddress,
 amsmath,amssymb,
 aps,longbibliography
]{revtex4-1}

\usepackage{graphicx}
\usepackage{dcolumn}
\usepackage{bm}
\usepackage{multirow}
\usepackage{upgreek}
\usepackage{braket}
\usepackage{bm}        
\usepackage{amssymb}   
\usepackage{amsmath}
\usepackage{subfigure}
\usepackage{epstopdf}
\usepackage{comment}
\usepackage{float}
\usepackage{xcolor}
\usepackage{xurl}
\usepackage{hyperref}
\usepackage{silence}
\begin{document}

\title{Neutronic Field--Effect Transistor}

\author{Dusan Sarenac}
\email{dusansar@buffalo.edu}
\affiliation{Department of Physics, University at Buffalo, State University of New York, Buffalo, New York 14260, USA}

\author{Charles W. Clark}
\affiliation{Joint Quantum Institute, National Institute of Standards and Technology and University of Maryland, College Park, Maryland 20742, USA}

\author{Owen Lailey} 
\affiliation{Institute for Quantum Computing, University of Waterloo,  Waterloo, ON, Canada, N2L3G1}
\affiliation{Department of Physics and Astronomy, University of Waterloo, Waterloo, ON, Canada, N2L3G1}

\author{Muhammad Arif}
\altaffiliation{Deceased}
\affiliation{National Institute of Standards and Technology, Gaithersburg, Maryland 20899, USA}

\author{Wangchun Chen}
\affiliation{National Institute of Standards and Technology, Gaithersburg, Maryland 20899, USA}

\author{David G. Cory}
\affiliation{Institute for Quantum Computing, University of Waterloo,  Waterloo, ON, Canada, N2L3G1}
\affiliation{Department of Chemistry, University of Waterloo, Waterloo, ON, Canada, N2L3G1}

\author{Michael G. Huber}
\affiliation{National Institute of Standards and Technology, Gaithersburg, Maryland 20899, USA}

\author{Joachim Nsofini} 
\affiliation{Institute for Quantum Computing, University of Waterloo,  Waterloo, ON, Canada, N2L3G1}
\affiliation{Department of Physics and Astronomy, University of Waterloo, Waterloo, ON, Canada, N2L3G1}

\author{Dmitry A. Pushin}
\email{dmitry.pushin@uwaterloo.ca}
\affiliation{Institute for Quantum Computing, University of Waterloo,  Waterloo, ON, Canada, N2L3G1}
\affiliation{Department of Physics and Astronomy, University of Waterloo, Waterloo, ON, Canada, N2L3G1}

\begin{abstract}
Field-effect electronic devices are in widespread use in communication, computing, data storage, and sensing applications. We demonstrate a neutronic field-effect transistor (NFET) that is analogous to the Datta--Das electron spin transistor and atomtronic devices. The NFET is based on a perfect-crystal Mach-Zehnder neutron interferometer with phase shifts applied to the spin and path degrees of freedom.
\end{abstract}
\maketitle

\section{Introduction}
Neutron interferometry is a powerful tool for investigating fundamental quantum physics and precision measurement~\cite{sam}. It also provides a platform for testing quantum computing concepts in a controlled, low-noise environment~\cite{dfs}, as well as performing weak measurements of neutron spin and path degrees of freedom~\cite{sponar2015weak,denkmayr2017experimental,lemmel2022quantifying}.

Here we view neutron interferometry from a spintronic perspective. We demonstrate a neutronic spin transistor that uses the same principle as the electron spin transistor proposed by Datta and Das~\cite{DattaDas1990,datta2018spintransistor}, that was subsequently realized by Koo, et al.~\cite{Koo2009}. Beyond simple switching applications, our device can also support universal quantum gate sets on the two-qubit neutron spin-path system.

\section{Field--effect devices}
\begin{figure}[H]
\center
\includegraphics[width=1\linewidth]{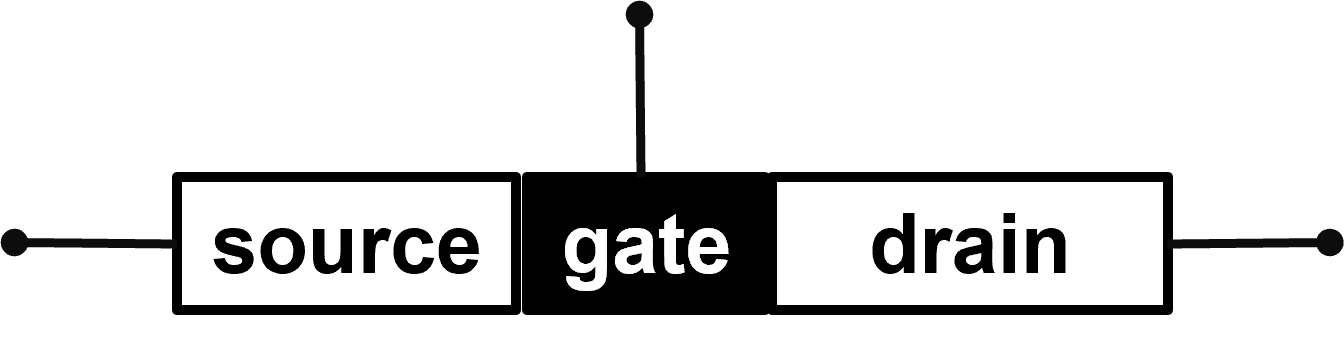}
\caption{General schematic of a field--effect device. A control field applied to the gate governs the current flow between source and drain.}
\label{fig:FET}
\end{figure}

Field--effect devices can be schematized by the generic Fig.~\ref{fig:FET}, where \textbf{\textsf{source}}  designates an input current, \textbf{\textsf{gate}} is an interface subject to a control field, and  \textbf{\textsf{drain}} is an output current. 

The most consequential early example is the field--effect transistor (FET), a concept traceable to the 1926 U.S. Patent Application filed by Julius E. Lilienfeld~\cite{Lilienfeld1926}. The FET \textit{gate} is a junction that controls the flow of electrical current  between an input source and an output drain. In 1952, William Shockley proposed a junction field--effect transistor that modulated current flow by electric--field control of a PN junction~\cite{Shockley1952}. The metal-oxide semiconductor field-effect transistor, MOSFET, introduced by Kahng and Atalla in 1960~\cite{dawon1963electric}, is now a fundamental building block of modern electronics. 

New gate physics can lead to new applications. For example, a major advance in field--effect transduction was heralded by the 2004 publication of Novoselov, \textit{et al.}~\cite{Novoselov2004}. There, the gate was made of a strip of effectively two--dimensional material, graphene, consisting of a few atomic layers of graphite. Graphene is a semimetal and can be rapidly switched between conducting and non-conducting states by a gate voltage swing of 10 -- 100 V. The advent of low--dimensional gate materials has led to new field--effect \textit{sensors}, as reviewed by Dai, Liu, and Wei~\cite{DLW2022}. 

\section{Spintronics}
The mid-1980's observation of ``giant magnetoresistance''~\cite{baibich1988gmr,binasch1989magnetoresistance} sparked revolutionary improvements in magnetic data storage. Manifested in thin multilayer films of alternating ferromagnetic materials with nonmagnetic spacers, it resulted in a 1000-fold increase in magnetic storage capacity. Recognized by the 2007 Nobel Prize in Physics, this development inspired visions of a future ``spintronics'' in which electron spin might supplement electric charge as a vehicle of logical bit representation in digital circuits. Influential early perspectives on spintronics are presented in~\cite{wolf2001spintronics} and~\cite{zutic2004spintronics}. More recent reviews emphasize data storage and sensing applications~\cite{Spintronics2024,jo2024spintronics,rimmler2025non,Editorial2024,antiferro}.

In 1990, Supriyo Datta and Biswajit Das proposed a field-effect electron {\em spin transistor}, as a candidate for a spintronic counterpart to the electronic FET~\cite{DattaDas1990}. This spintronic concept was realized in a working device 19 years later~\cite{Koo2009}. Meanwhile, it inspired exploration of potential uses in quantum computing~\cite{awschalom2002spintronics}; implementation in atomtronic devices~\cite{vaishnav2008spin, madasu2022datta, RevModPhys.94.041001, amico2021roadmap, PhysRevA.109.053319, ramsey, 7kb2-118m, Kwek2026}; and the neutron spin transistor reported here.

\begin{figure}[H]
\center
\includegraphics[width=1\linewidth]{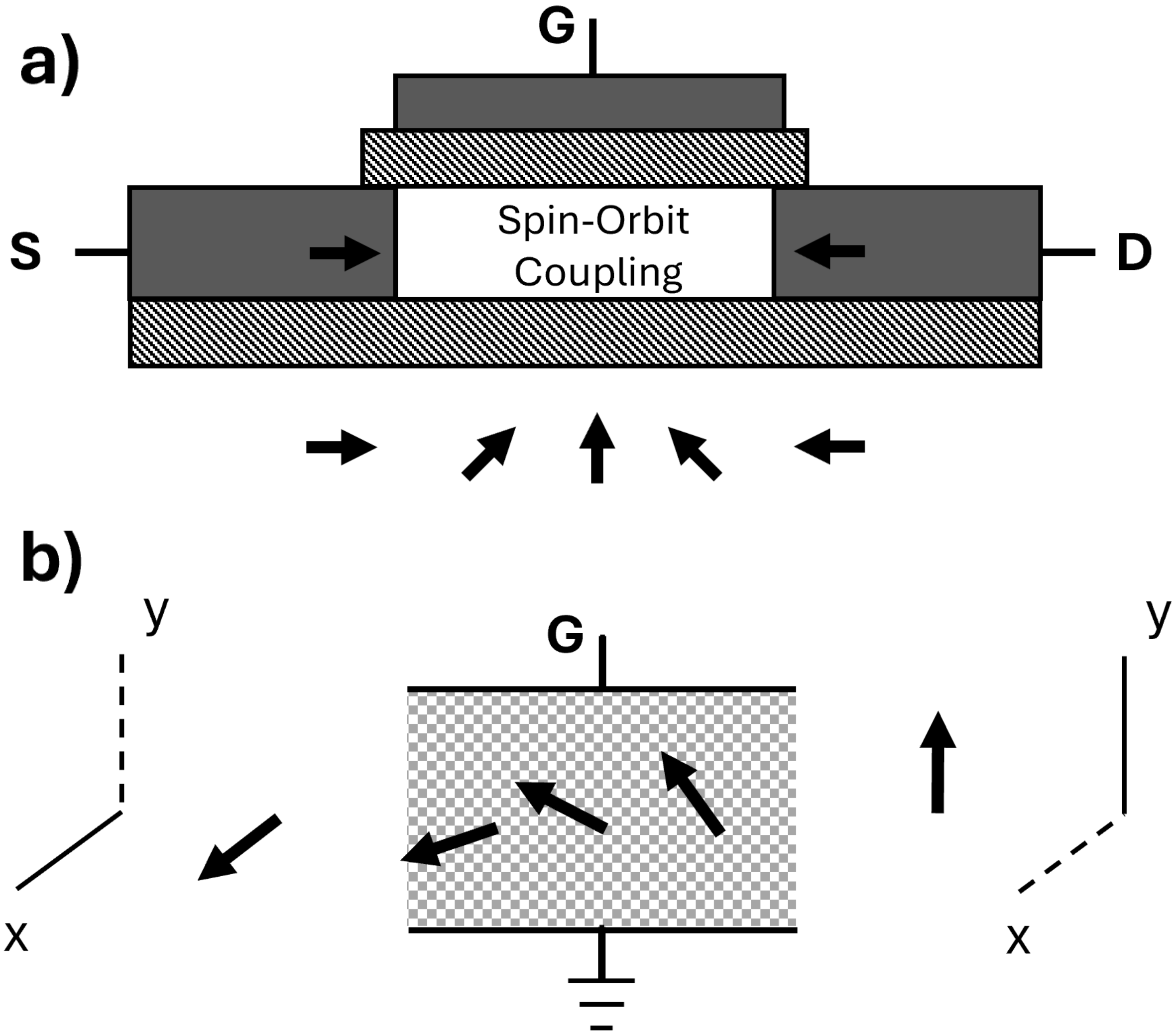}
\caption{The Datta-Das transistor and its optoelectronic analog, the Pockels cell. Both of these are polarization--rotation devices. \textsf{\textbf{a)}} The source, \textsf{\textbf{S}}, is a magnetized source of electrons that are spin--polarized in the direction of electron current flow; the drain, \textsf{\textbf{D}}, is magnetized so that only electrons with opposite spin are admitted.  Between \textsf{\textbf{S}} and \textsf{\textbf{D}} there is a semiconductor medium that supports a voltage--controlled spin--orbit interaction, e.g. of the Rashba type demonstrated by Grundler~\cite{Grundler2000}. When the voltage of the gate, \textbf{\textsf{G}} attains a target value, the spin--orbit coupling interaction rotates the electron spin by 180\textdegree $ $  during the electron transit through the indicated region. \textsf{\textbf{b)}} Schematic of a Pockels cell, in which the birefringence of transparent crystal can be tuned by the voltage applied to gate \textbf{\textsf{G}} to rotate a horizontally--polarized photon into vertical polarization (\cite{waynant2000electro}). Figure adapted from Ref. \cite{datta2018spintransistor}.}
\label{fig:Pockels}
\end{figure}

Fig.~\ref{fig:Pockels} shows that the Datta--Das spin transistor is fundamentally a spin--rotation device.  As noted by Datta~\cite{datta2018spintransistor}, ``a typical FET has an OFF/ON resistance [extinction] ratio of about $10^5$. By contrast, the spin transistor can at best reduce the current a little bit, partly because polarizers and analyzers are far from ideal and partly because scattering processes in solids can randomize spins quite strongly . . . Alternative designs have been proposed to overcome this limitation.''

Polarization extinction ratios of 10$^5$ are readily obtainable with photons~\cite{PhysRevApplied.19.064082}, and indeed MOSFET technology is now being deployed in electronic switches~\cite{powerFET}. In contrast to the centuries over which light polarization technology evolved, sources and detectors of spin-polarized electrons date back only to the 1960s~\cite{10.1063/1.2888095}. Thus, future developments may bring the Datta--Das transistor closer to practicality.

Meantime, the spin polarization rejection ratios attained for neutron beams routinely exceed 1000, and in dedicated efforts they have attained about 10$^{12}$~\cite{Treimer_2020}.

Here we apply these ideas to realize a neutron spin transistor. This device achieves matter-wave amplitude modulation through quantum phase control mediated by a neutron interferometer.

\section{The perfect-crystal neutron interferometer}
Since its realization in 1974~\cite{rauch1974}, the perfect-crystal neutron interferometer (NI) has found numerous applications in studies of materials and fundamental physics~\cite{sam,fourpi,colella1975observation}. Recent advances in the decoupling of the NI from the environment~\cite{dfs, saggu2016decoupling} have allowed the implementation of demanding and sensitive experiments~\cite{Dima2015,chameleon,sarenac2016holography}. Here we demonstrate a unique function of the NI which has diverse integrated applications.

The oscillation of spin polarization at the output of an interferometer due to simultaneously applied phase shifts on the path and spin degrees of freedom was first noted and observed by Badurek \textit{et al.} in $1976$~\cite{badurek1976phase}. They achieved a polarization of $20\%$. In $2001$ the same concept was proposed for electron spin polarizer devices~\cite{kiselev2001t}. In the field of spintronics this is a possible solution to the major problem of having a controlled source of spin polarized electrons. Later on this approach was generalized to non dispersive degrees of freedom which would yield higher magnitude correlations~\cite{radu}.

Here we demonstrate a neutron spin transistor with a maximum output polarization of $P = 77 \pm 2~\%$, corresponding to a spin extinction ratio $(1+P)/(1-P) \approx 8{:}1$, limited by the $83~\%$ contrast of the interferometer.

\section{Design and Theory}
\begin{figure}
\center
\includegraphics[width=1\linewidth]{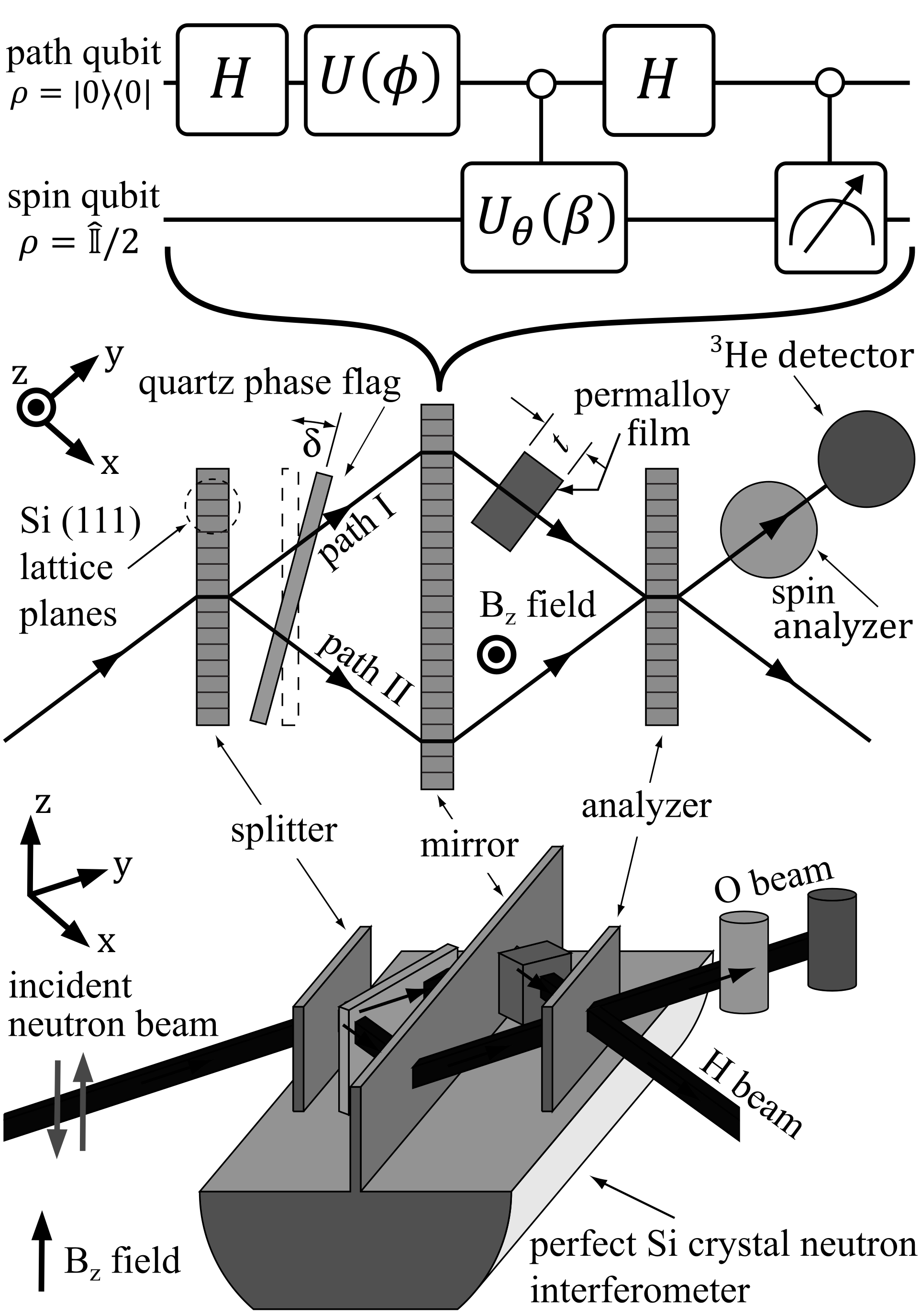}
\caption{Schematic of a standard perfect-crystal neutron interferometer, adapted from Ref.~\cite{schoen2003precision}.  The lower portion is a drawing of the device; the middle portion is a functional schematic, from a viewpoint above the $x-y$ plane defined by the neutron beams; the upper portion is the corresponding quantum gate description of the device. The incident neutron beam is unpolarized. A uniform guide field $B_z$ is applied along $z$ over the full length of the interferometer and the spin analyzer, defining the spin quantization axis. In the circuit, $H$ is the Hadamard gate representing the beam splitting at the first and third blades, $U(\phi)$ is the phase flag rotation gate, and the controlled $U_\theta(\beta)$ is the spin rotation applied only in path~I (path state $|0\rangle$). The final spin measurement, conditioned on the path, corresponds to analyzing the spin of the O beam.}
 \label{fig:NI}
\end{figure}

Our neutron spin transistor is built upon a perfect--crystal neutron interferometer.  Presuming no previous familiarity by the reader with that device, we summarize its basic operations with reference to Fig.~\ref{fig:NI}. 

The lower portion of Fig.~\ref{fig:NI} is a drawing of the interferometer, the middle portion is a diagram of its operation, as seen from above, and the upper portion is the corresponding quantum circuit description of the device. 

The interferometer chassis is made from a cylindrical boule of high--purity single--crystal silicon. The upper half of that boule was reduced by subtractive milling to a structure of three blades standing in the atomic registry of the original crystal. As labeled, those blades serve the functions of beam--splitter, mirror, and analyzer of an optical Mach--Zehnder interferometer~\cite{Mach1892, Zehnder1891}, except that the ``mirrors'' are Bragg--diffracting crystals that transmit part of their incident neutron beams out of the interferometer.

The left--most blade serves as a beam splitter: a fraction of the incident neutron beam is transmitted through it, another is diffracted at the Bragg angle. The middle blade serves as a mirror: part of the beam incident on it is diffracted at the Bragg angle; the other part (not shown here) is transmitted and lost to the interferometer. The rightmost blade coherently recombines the two paths and separates them into the forward--diffracted (O) beam and the diffracted (H) beam. We post-select on only those neutrons that reach the spin analyzer in the O-beam. 

In the application shown here, the quartz phase flag induces a differential quantum phase shift ($\phi$) between neutron paths {I} and {II} that depends upon the angle $\delta$. This is due to the difference between the neutron indices of refraction of air and quartz.

In addition to the path phase $\phi$, a spin rotation $U_\theta(\beta)$ is applied in path~I only. It is a rotation by $\beta$ about the axis $\hat{n} = (0, \sin\theta, \cos\theta)$, which lies in the $y$--$z$ plane of the Bloch sphere at an angle $\theta$ from $z$:
\begin{equation}
U_\theta(\beta) = e^{-i\beta\,\hat{n}\cdot\boldsymbol{\sigma}/2},
\end{equation}
where $\boldsymbol{\sigma} = (\sigma_x, \sigma_y, \sigma_z)$ are the Pauli matrices. The spin components parallel and antiparallel to $\hat{n}$ are eigenstates of $U_\theta(\beta)$ and acquire opposite phases, $\mp\beta/2$, in path~I. This spin-dependence can be used to convert an unpolarized input into a polarized output, with $\phi$, $\beta$, and $\theta$ acting as the gate parameters.

In our implementation, $U_\theta(\beta)$ is applied by a $10~\mu$m-thick permalloy film~\cite{wood2014quantum}. The two paths are separated by several centimeters, so the film can be placed in path~I alone. Permalloy is a soft ferromagnet; the film is magnetized in its plane along $z$, so $\theta = 0$. The neutron spin precession angle $\beta$ is proportional to the neutron's path length through the film; rotating the film about $z$ lengthens this path and so tunes $\beta$.

The third blade coherently recombines the two paths, outputting the O beam and the H beam. These two outputs are projections onto the eigenstates of the path qubit, and we post-select on the O beam. For an unpolarized incident beam, its intensity is
\begin{equation}
I_O = \frac{1}{2}\left[1 + \cos\tfrac{\beta}{2}\cos\phi\right],
\label{Obeam}
\end{equation}
and its polarization vector is
\begin{equation}
\mathbf{P} =
\frac{\sin\frac{\beta}{2}\sin\phi}{1+\cos\frac{\beta}{2}\cos\phi}
\begin{bmatrix}
0 \\
\sin\theta \\
\cos\theta
\end{bmatrix},
\label{eq:pol}
\end{equation}
so the spin state of the O beam is
\begin{equation}
\rho_O = \tfrac{1}{2}\left(\mathbb{I} + \mathbf{P}\cdot\boldsymbol{\sigma}\right).
\end{equation}

In the configuration used here ($\beta = \pi$, $\theta = 0$), the O-beam intensity is constant, $I_O = 1/2$, and its spin state is $\rho_O = \tfrac{1}{2}(\mathbb{I} + \sin\phi\,\sigma_z)$: varying $\phi$ changes the polarization along $z$ without changing the intensity. We measure $P_z = (N_+ - N_-)/(N_+ + N_-)$ from the counts transmitted by a $+z$ spin analyzer with a spin flipper off ($N_+$) and on ($N_-$).

Experimentally, the maximum achievable polarization is limited by the contrast of the interferometer. The contrast is a figure of merit of the entire NI setup quality and it is defined by the intensity at the O or H detector as
\begin{align}
C=\frac{I_{max}-I_{min}}{I_{max}+I_{min}}
\end{align}
Because the NI blades are not $50:50$ beam splitters the contrast at the H-detector can not reach 1. While the contrast at the O-detector is in principle 1. 

\begin{figure}
\centering\includegraphics[width=1\linewidth]{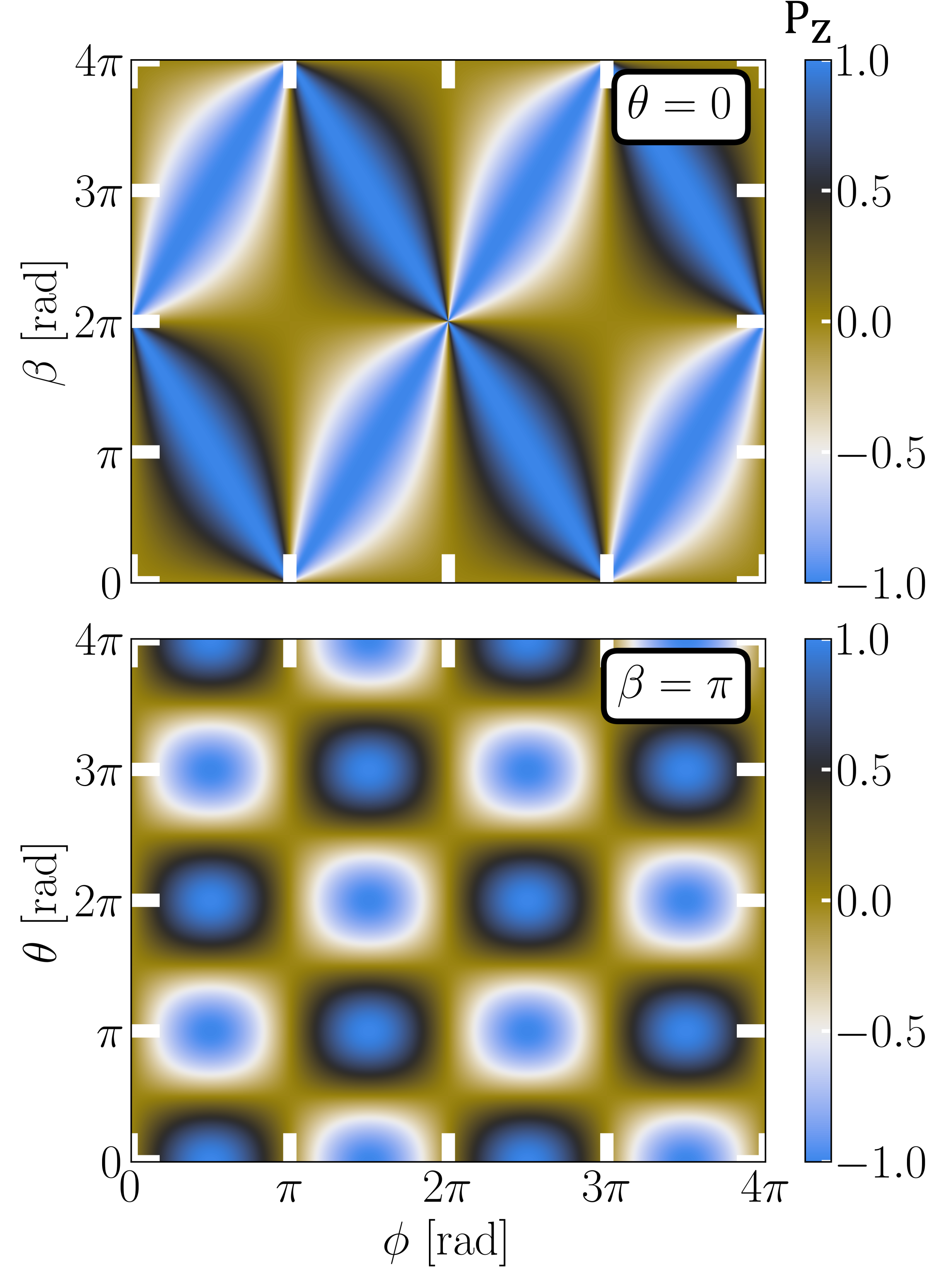}
    \caption{Calculated O-beam polarization along the $z$ direction following a spin rotation $U_\theta(\beta)$ and the application of a phase flag $\phi$ in one interferometer path, according to Eq.~\ref{eq:pol}. The upper panel shows the polarization as a function of $\phi$ and $\beta$ for $\theta=0$. The lower panel shows the polarization as a function of $\phi$ and $\theta$ for $\beta=\pi$.}
    \label{fig:sims}
\end{figure}

Figure~\ref{fig:sims} illustrates the transfer characteristics of the neutron spin transistor. The output polarization is jointly controlled by the interferometric phase flag $\phi$, the spin rotation magnitude $\beta$, and the rotation axis $\theta$. The upper panel ($\theta=0$) shows that the O-beam polarization along the spin quantization axis can be continuously tuned between positive and negative values by varying $\phi$ and $\beta$, corresponding to switching between spin-up and spin-down output states. The lower panel ($\beta=\pi$) shows the corresponding dependence on $\phi$ and $\theta$, demonstrating that the output polarization can be continuously tuned through the combined control of the interferometric phase and spin rotation. 

The Datta--Das drain current is the fraction of electrons transmitted by the spin-selective drain, $\tfrac{1}{2}(1+\cos\Delta)$, where $\Delta$ is the gate-controlled precession angle. In our device the spin analyzer plays the role of the drain: it transmits a fraction $\tfrac{1}{2}(1+P_z)$ of the O beam, converting its polarization into intensity. Our device reproduces the Datta--Das transfer characteristic in two ways. At $\phi = \pi/2$ and $\beta = \pi$, Eq.~\ref{eq:pol} gives $\mathbf{P} = (0, \sin\theta, \cos\theta)$: the O beam is fully polarized along the rotation axis, $P_z = \cos\theta$, and the analyzer transmits $\tfrac{1}{2}(1+\cos\theta)$, so $\theta$ would play the role of $\Delta$. At $\beta = \pi$ and $\theta = 0$, $P_z = \sin\phi$, and the analyzer transmits $\tfrac{1}{2}[1 + \sin\phi]$, with the nuclear phase $\phi$ acting as the gate. In this case the polarization along $z$ varies continuously between $\pm 1$ and vanishes at $\phi = 0$.

Notably, our device is designed to accept an unpolarized beam. The polarization at the drain is produced by spin-dependent interference between the two paths, so the interferometer serves as both the spin-polarizing source and the gate. 

\section{Implementation and Results}
The measurements were performed at the NIOF beamline at the National Center for Neutron Research (NCNR) at the National Institute of Standards and Technology~\cite{pushin2015neutron, shahi2016new}. The NIOF beamline is dedicated to neutron interferometry experiments. The neutrons delivered to the interferometer have a de Broglie wavelength of $\lambda_\mathrm{dB}=0.271$~nm.

The neutron interferometer was measured to have an O-beam contrast of $83~\%$, while the incident beam was unpolarized. A $10~\mu$m-thick permalloy film with a saturation field of $0.9~$T was placed in one arm of the interferometer. The O-beam contrast was measured as a function of the film rotation to identify the orientation corresponding to a magnetic phase shift of $\beta=\pi$. At this condition, Eq.~\ref{Obeam} predicts zero O-beam contrast. The film was then fixed at this orientation, and the O-beam polarization was measured as a function of the phase-flag rotation.

\begin{figure}[!b]
    \centering\includegraphics[width=1\linewidth]{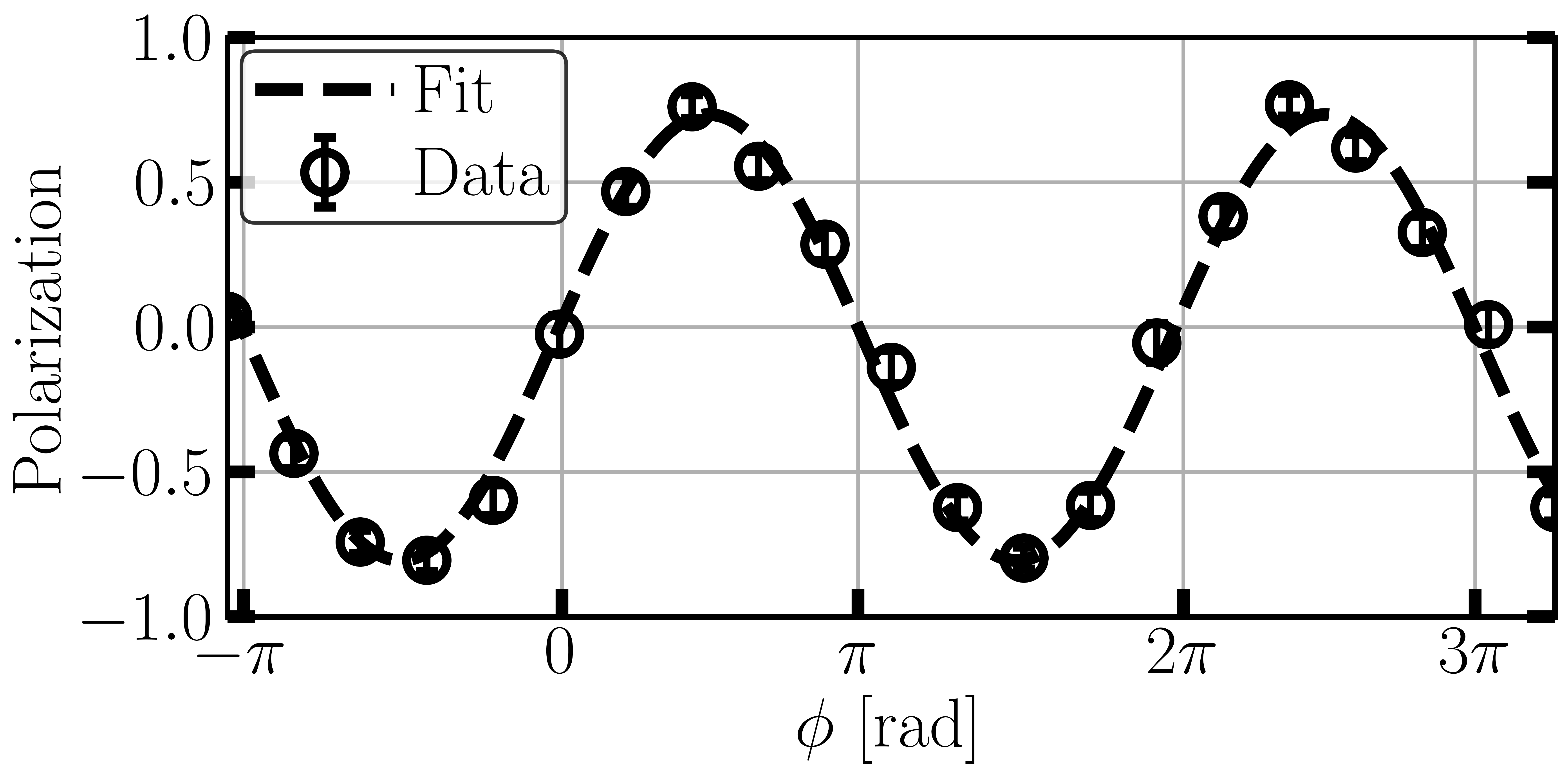}
    \caption{Measured O-beam polarization $P_z$ as a function of $\phi$ at $\beta = \pi$, $\theta = 0$. This illustrates direct control of the spin state in the forward diffracted path at the output of the interferometer by tuning $\phi$ and $\beta$. The uncertainties, which are smaller than the data markers shown, are purely statistical.}
    \label{fig:Opol}
\end{figure}

Figure~\ref{fig:Opol} shows the measured O-beam polarization as a function of the phase-shift $\phi$. The maximum measured polarization was $77~\%\pm2~\%$, consistent with the $83~\%$ interferometer contrast setting the upper limit on the achievable output polarization. The observed variation of polarization with phase-flag angle demonstrates experimental control of the output spin state through the interferometric phase. For this neutronic field-effect transistor, the incident unpolarized neutron beam serves as the source, the nuclear and magnetic phase shifts act as gate controls, and the O beam, passed through the spin analyzer, serves as the drain. The gates set the spin state of the O beam, and the analyzer converts it into a transmitted intensity, the analog of the drain current. At $\beta=\pi$, varying the phase $\phi$ continuously changes the polarization of the O-beam, allowing the output to be tuned between the two spin states.

This mapping also provides a route to single-shot spin readout~\cite{elzerman2004single, morello2010single, awschalom2013quantum}. In the neutron implementation, each $^3$He absorption event corresponds intrinsically to the detection of an individual neutron, while the low neutron flux ensures that the mean occupancy of the interferometer is much less than one neutron~\cite{pushin2015neutron}. Thus, while one neutron is being measured, the next neutron has typically not yet reached the interferometer. When the phase and magnetic rotation are chosen such that the O-beam corresponds predominantly to one spin state, the output path of an individual detected neutron therefore provides a single-neutron spin measurement, with a fidelity limited by the measured output polarization.

\section{Conclusion and Discussion}

In this work, we have demonstrated a neutron spin transistor in which a magnetic spin rotation and a nuclear phase flag in a neutron interferometer control the polarization of the output beam from an unpolarized input. Paired with analyzers, this polarization modulation shapes intensity profiles, mirroring the gate control of electronic field effect transistors. The ability to access and tune neutron polarization, alongside beam splitting, paves the way for applications in neutron scattering, quantum information, and materials analysis. Neutrons, with their charge-free spin, offer a fresh perspective on spintronic applications. 

\section*{Acknowledgments}
This work was supported by the Canadian Excellence Research Chairs (CERC) program, the Natural Sciences and Engineering Research Council of Canada (NSERC) Discovery program, the NSERC Canada Graduate Scholarships programs (CGS-M and PGS-D), the Canada  First  Research  Excellence  Fund  (CFREF), and the US Department of Energy, Office of Nuclear Physics, under Interagency Agreement 89243019SSC000025. Unless otherwise noted, work of the National Institute of Standards and Technology was funded solely by the United States Government.

\bibliographystyle{ieeetr}
\bibliography{SSM}

@article{Spintronics2024,author={Incorvia, Jean A. C. 
and Xiao,T. P. and Zogbi,Nicholas and Naeemi,Azad and Adelmann,Christoph and Catthoor,Francky and Tahoori,Mehdi and Casanova,Fè and Becherer,Markus and Prenat,Guillaume and Couet,Sebastien},
year={2024},
month={11},
title={Spintronics for achieving system-level energy-efficient logic},
journal={Nature Reviews.Electrical Engineering},
volume={1},
number={11},
pages={700-713}
}

@article{antiferro, author = {A. Dal Din, O. J. Amin, P. Wadley and K. W. Edmonds},
title={Antiferromagnetic spintronics and beyond},
year={2024},
journal = {npj Spintronics},
volume = {2}, 
number = {25}
}

@article{jo2024spintronics,
  title={Spintronics meets orbitronics: Emergence of orbital angular momentum in solids},
  author={Jo, Daegeun and Go, Dongwook and Choi, Gyung-Min and Lee, Hyun-Woo},
  journal={npj Spintronics},
  volume={2},
  number={1},
  pages={19},
  year={2024},
  publisher={Nature Publishing Group UK, London}
}

@article{sarenac2016holography,
  title={Holography with a neutron interferometer},
  author={Sarenac, Dusan and Huber, Michael G and Heacock, Benjamin and Arif, Muhammad and Clark, Charles W and Cory, David G and Shahi, Chandra B and Pushin, Dmitry A},
  journal={Optics express},
  volume={24},
  number={20},
  pages={22528--22535},
  year={2016},
  publisher={Optical Society of America}
}

@article{rimmler2025non,
  title={Non-collinear antiferromagnetic spintronics},
  author={Rimmler, Berthold H and Pal, Banabir and Parkin, Stuart SP},
  journal={Nature Reviews Materials},
  volume={10},
  number={2},
  pages={109--127},
  year={2025},
  publisher={Nature Publishing Group UK, London}
}

@article{Editorial2024,
title={Spintronics for ultra-low-power circuits and systems},
journal={Nature Reviews Electrical Engineering},
volume={1},
pages={691},
year={2024}
}

@article{DLW2022,
author = {Dai, Changhao and Liu, Yunqi and Wei, Dacheng},
title = {Two-Dimensional Field-Effect Transistor Sensors: The Road toward Commercialization},
journal = {Chemical Reviews},
volume = {122},
number = {11},
pages = {10319-10392},
year = {2022},
doi = {10.1021/acs.chemrev.1c00924},
note ={PMID: 35412802},
URL = { https://doi.org/10.1021/acs.chemrev.1c00924},
eprint = {https://doi.org/10.1021/acs.chemrev.1c00924}
}

@article{Novoselov2004,
  author = {Novoselov, K. S. and Geim, A. K. and Morozov, S. V. and Jiang, D. and Zhang, Y. and Dubonos, S. V. and Grigorieva, I. V. and Firsov, A. A.},
  title = {Electric Field Effect in Atomically Thin Carbon Films},
  journal = {Science},
  volume = {306},
  number = {5696},
  pages = {666--669},
  year = {2004},
  month = {October},
  doi = {10.1126/science.1102896},
  url = {https://doi.org/10.1126/science.1102896}
}

@misc{Lilienfeld1926,
  author = {Lilienfeld, Julius Edgar},
  title = {Method and apparatus for controlling electric currents},
  year = {1930},
  month = {January},
  howpublished = {U.S. Patent No. 1,745,175},
  note = {Filed October 8, 1926},
}

@article{Shockley1952,
  author = {Shockley, William},
  title = {A Unipolar Field-Effect Transistor},
  year = {1952},
  journal = {Proceedings of the I. R. E.},
  volume = {40},
  pages = {1365--1376}
}

@misc{dawon1963electric,
  title={Electric field controlled semiconductor device},
  author={Kahng, Dawon},
  year={1963},
  month=aug # "~27",
  publisher={Google Patents},
  note={US Patent 3,102,230}
}

@article{baibich1988gmr,
  author = {Baibich, M. N. and Broto, J. M. and Fert, A. and Nguyen Van Dau, F. and Petroff, F. and Etienne, P. and Creuzet, G. and Friederich, A. and Chazelas, J.},
  title = {Giant Magnetoresistance of (001){Fe}/(001){Cr} Magnetic Superlattices},
  journal = {Physical Review Letters},
  volume = {61},
  number = {21},
  pages = {2472--2475},
  year = {1988},
  month = {November},
  doi = {10.1103/PhysRevLett.61.2472},
  publisher = {American Physical Society}
}

@article{binasch1989magnetoresistance,
  author = {Binasch, G. and Grünberg, P. and Saurenbach, F. and Zinn, W.},
  title = {Enhanced Magnetoresistance in Layered Magnetic Structures with Antiferromagnetic Interlayer Exchange},
  journal = {Physical Review B},
  volume = {39},
  number = {7},
  pages = {4828--4830},
  year = {1989},
  month = {March},
  doi = {10.1103/PhysRevB.39.4828},
  publisher = {American Physical Society}
}

@article{wolf2001spintronics,
  author = {Wolf, S. A. and Awschalom, D. D. and Buhrman, R. A. and Daughton, J. M. and von Molnár, S. and Roukes, M. L. and Chtchelkanova, A. Y. and Treger, D. M.},
  title = {Spintronics: A Spin-Based Electronics Vision for the Future},
  journal = {Science},
  volume = {294},
  number = {5546},
  pages = {1488--1495},
  year = {2001},
  month = {November},
  doi = {10.1126/science.1065389},
  publisher = {American Association for the Advancement of Science}
}

@article{zutic2004spintronics,
  author = {Žutić, Igor and Fabian, Jaroslav and Das Sarma, S.},
  title = {Spintronics: Fundamentals and Applications},
  journal = {Reviews of Modern Physics},
  volume = {76},
  number = {2},
  pages = {323--410},
  year = {2004},
  month = {April},
  doi = {10.1103/RevModPhys.76.323},
  publisher = {American Physical Society}
}

@article{DattaDas1990,
  author = {Datta, Supriyo and Das, Biswajit},
  title = {Electronic Analog of the Electro-optic Modulator},
  journal = {Applied Physics Letters},
  volume = {56},
  number = {7},
  pages = {665--667},
  year = {1990},
  month = {February},
  doi = {10.1063/1.102730},
  publisher = {American Institute of Physics}
}

@article{datta2018spintransistor,
  author = {Datta, Supriyo},
  title = {How We Proposed the Spin Transistor},
  journal = {Nature Electronics},
  volume = {1},
  number = {11},
  pages = {604},
  year = {2018},
  month = {November},
  doi = {10.1038/s41928-018-0163-4},
  publisher = {Nature Publishing Group}
}

@incollection{awschalom2002spintronics,
  author = {Awschalom, D. D. and Loss, D. and Samarth, N.},
  title = {Semiconductor Spintronics and Quantum Computation},
  booktitle = {Semiconductor Spintronics and Quantum Computation},
  editor = {Awschalom, D. D. and Loss, D. and Samarth, N.},
  pages = {1--36},
  year = {2002},
  publisher = {Springer},
  doi = {10.1007/978-3-662-05003-3_1}
}

@article{vaishnav2008spin,
  author = {Vaishnav, J. Y. and Ruseckas, Julius and Clark, Charles W. and Juzeliūnas, Gediminas},
  title = {Spin Field Effect Transistors with Ultracold Atoms},
  journal = {Physical Review Letters},
  volume = {101},
  number = {26},
  pages = {265302},
  year = {2008},
  month = {December},
  doi = {10.1103/PhysRevLett.101.265302},
  publisher = {American Physical Society}
}

@article{madasu2022datta,
  author = {Madasu, Chetan Sriram and Hasan, Mehedi and Kwong, Chang Chi and Wilkowski, David},
  title = {Datta-Das Transistor for Atomtronic Circuits Using Artificial Gauge Fields},
  journal = {Physical Review Research},
  volume = {4},
  number = {3},
  pages = {033180},
  year = {2022},
  month = {September},
  doi = {10.1103/PhysRevResearch.4.033180},
  publisher = {American Physical Society}
}

@article{Grundler2000,
  author = {Grundler, D.},
  title = {Large Rashba Splitting in InAs Quantum Wells due to Electron Wave Function Penetration into the Barrier Layers},
  journal = {Physical Review Letters},
  volume = {84},
  number = {26},
  pages = {6074--6077},
  year = {2000},
  month = {jun},
  publisher = {American Physical Society},
  doi = {10.1103/PhysRevLett.84.6074},
  url = {https://journals.aps.org/prl/pdf/10.1103/PhysRevLett.84.6074}
}

@article{MOSFET,
author = {Yadlapalli, Ravindranath Tagore and Narasipuram, Rajanand Patnaik and Kotapati, Anuradha},
title = {An overview of energy efficient solid state LED driver topologies},
journal = {International Journal of Energy Research},
volume = {44},
number = {2},
pages = {612-630},
doi = {https://doi.org/10.1002/er.4924},
url = {https://onlinelibrary.wiley.com/doi/abs/10.1002/er.4924},
eprint = {https://onlinelibrary.wiley.com/doi/pdf/10.1002/er.4924},
year = {2020}
}

@article{Dima2015,
  title = {Controlling neutron orbital angular momentum},
  author = {C. W. Clark and R. Barankov and M. G. Huber and D. G. Cory and D. A. Pushin},
  journal = {Nature},
  volume = {525},
  issue = {7570},
  pages = {504--506},
  numpages = {0},
  year = {2015},
  month = {Sep},
  publisher = {Nature Publishing Group},
  doi = {http://dx.doi.org/10.1038/525462a},
  url = {http://www.nature.com/nature/journal/v525/n7570/abs/525462a.html}
  }

@article{badurek1976phase,
  title={Phase-shift and spin-rotation phenomena in neutron interferometry},
  author={Badurek, Gerald and Rauch, Helmut and Zeilinger, Anton and Bauspiess, W and Bonse, U},
  journal={Physical Review D},
  volume={14},
  number={5},
  pages={1177},
  year={1976},
  publisher={APS}
}

@article{kiselev2001t,
  title={T-shaped ballistic spin filter},
  author={Kiselev, Andrey A and Kim, Ki Wook},
  journal={Applied Physics Letters},
  volume={78},
  number={6},
  pages={775--777},
  year={2001},
  publisher={AIP Publishing}
}

@article{rauch1974,
  title={Test of a single crystal neutron interferometer},
  author={Rauch, H and Treimer, W and Bonse, U},
  journal={Physics Letters A},
  volume={47},
  number={5},
  pages={369--371},
  year={1974},
  publisher={Elsevier}
}

@book{sam,
  title={Neutron Interferometry: Lessons in Experimental Quantum Mechanics, Wave-particle Duality, and Entanglement},
  author={Rauch, Helmut and Werner, Samuel A},
  volume={12},
  year={2015},
  publisher={Oxford University Press}
}

@book{waynant2000electro,
  title = {Electro-Optics Handbook},
  editor = {Ronald W. Waynant and Marwood N. Ediger},
  edition = {2nd},
  publisher = {McGraw-Hill},
  year = {2000},
  address = {New York},
  isbn = {0-07-068716-1},
  note = {McGraw-Hill Optical and Electro-Optical Engineering Series}
}

@article{colella1975observation,
  title={Observation of gravitationally induced quantum interference},
  author={Colella, Roberto and Overhauser, Albert W and Werner, Samuel A},
  journal={Physical Review Letters},
  volume={34},
  number={23},
  pages={1472},
  year={1975},
  publisher={APS}
}

@article{fourpi,
  title={Verification of coherent spinor rotation of fermions},
  author={Rauch, Helmut and Zeilinger, Anton and Badurek, Gerald and Wilfing, A and Bauspiess, W and Bonse, U},
  journal={Physics Letters A},
  volume={54},
  number={6},
  pages={425--427},
  year={1975},
  publisher={Elsevier}
}

@article{dfs,
  title={Experimental realization of decoherence-free subspace in neutron interferometry},
  author={Pushin, DA and Huber, MG and Arif, M and Cory, DG},
  journal={Physical Review Letters},
  volume={107},
  number={15},
  pages={150401},
  year={2011},
  publisher={APS}
}

@article{radu,
  title={Mesoscopic Stern-Gerlach device to polarize spin currents},
  author={Ionicioiu, Radu and D’Amico, Irene},
  journal={Physical Review B},
  volume={67},
  number={4},
  pages={041307},
  year={2003},
  publisher={APS}
}

@article{chameleon,
  title = {Neutron limit on the strongly-coupled chameleon field},
  author = {Li, K. and Arif, M. and Cory, D. G. and Haun, R. and Heacock, B. and Huber, M. G. and Nsofini, J. and Pushin, D. A. and Saggu, P. and Sarenac, D. and Shahi, C. B. and Skavysh, V. and Snow, W. M. and Young, A. R.},
  collaboration = {The INDEX Collaboration},
  journal = {Phys. Rev. D},
  volume = {93},
  issue = {6},
  pages = {062001},
  numpages = {9},
  year = {2016},
  month = {Mar},
  publisher = {American Physical Society},
  doi = {10.1103/PhysRevD.93.062001},
  url = {http://link.aps.org/doi/10.1103/PhysRevD.93.062001}
}

@article{Koo2009,
  author = {Koo, Hyun Cheol and Kwon, Jae Hyun and Eom, Jonghwa and Chang, Joonyeon and Han, Suk Hee and Johnson, Mark},
  title = {Control of Spin Precession in a Spin-Injected Field Effect Transistor},
  journal = {Science},
  volume = {325},
  number = {5947},
  pages = {1515--1518},
  year = {2009},
  month = sep,
  issn = {0036-8075},
  doi = {10.1126/science.1173667},
  url = {https://www.science.org/doi/full/10.1126/science.1173667}
}

@article{ramsey,
  title = {Geometric Ramsey interferometry with a tripod scheme},
  author = {Madasu, Chetan Sriram and Rathod, Ketan Damji and Kwong, Chang Chi and Wilkowski, David},
  journal = {Phys. Rev. Appl.},
  volume = {21},
  issue = {5},
  pages = {L051001},
  numpages = {6},
  year = {2024},
  month = {May},
  publisher = {American Physical Society},
  doi = {10.1103/PhysRevApplied.21.L051001},
  url = {https://link.aps.org/doi/10.1103/PhysRevApplied.21.L051001}
}

@article{Kwek2026,
    author = {Hue, Jun Hao and Trappe, Martin-Isbjörn and Grochowski, Piotr T. and Lau, Jonathan and Kwek, Leong-Chuan},
    title = {A passive atomtronics filter for Fermi gases},
    journal = {AVS Quantum Science},
    volume = {8},
    number = {3},
    pages = {036801},
    year = {2026},
    month = {07},
    issn = {2639-0213},
    doi = {10.1116/5.0311396},
    url = {https://doi.org/10.1116/5.0311396}
}

@article{PhysRevApplied.19.064082,
  title = {Cross-Polarization-Extinction Enhancement and Spin-Orbit Coupling of Light for Quantum-Dot Cavity Quantum Electrodynamics Spectroscopy},
  author = {Steindl, P. and Frey, J.A. and Norman, J. and Bowers, J.E. and Bouwmeester, D. and L\"offler, W.},
  journal = {Phys. Rev. Appl.},
  volume = {19},
  issue = {6},
  pages = {064082},
  numpages = {7},
  year = {2023},
  month = {Jun},
  publisher = {American Physical Society},
  doi = {10.1103/PhysRevApplied.19.064082},
  url = {https://link.aps.org/doi/10.1103/PhysRevApplied.19.064082}
}

@article{10.1063/1.2888095,
    author = {Sinclair, Charles K.},
    title = {The Development of Polarized Electron Sources ‐ Physics and History},
    journal = {AIP Conference Proceedings},
    volume = {980},
    number = {1},
    pages = {25-36},
    year = {2008},
    month = {02},
    issn = {0094-243X},
    doi = {10.1063/1.2888095},
    url = {https://doi.org/10.1063/1.2888095}
}

@article{Treimer_2020,
doi = {10.1088/1361-6501/ab8cff},
url = {https://doi.org/10.1088/1361-6501/ab8cff},
year = {2020},
month = {aug},
publisher = {IOP Publishing},
volume = {31},
number = {11},
pages = {115017},
author = {Treimer, Wolfgang and Höppner, Henning},
title = {Realizing a (nearly) 100\% neutron beam polarization},
journal = {Measurement Science and Technology}
}

@article{amico2021roadmap,
  title={Roadmap on Atomtronics: State of the art and perspective},
  author={Amico, Luigi and Boshier, Malcolm and Birkl, Gerhard and Minguzzi, Anna and Miniatura, Christian and Kwek, L-C and Aghamalyan, Davit and Ahufinger, Veronica and Anderson, Dana and Andrei, Natan and others},
  journal={AVS Quantum Science},
  volume={3},
  number={3},
  year={2021},
  publisher={AIP Publishing}
}

@article{7kb2-118m,
  title = {Proposals for realizing a Josephson diode in atomtronic circuits},
  author = {Pradhan, Nalinikanta and Kanamoto, Rina and Bhattacharya, M. and Mishra, Pankaj Kumar},
  journal = {Phys. Rev. A},
  volume = {113},
  issue = {6},
  pages = {L061502},
  numpages = {7},
  year = {2026},
  month = {Jun},
  publisher = {American Physical Society},
  doi = {10.1103/7kb2-118m},
  url = {https://link.aps.org/doi/10.1103/7kb2-118m}
}

@article{PhysRevA.109.053319,
  title = {Magnetically generated spin-orbit coupling for ultracold atoms with slowly varying periodic driving},
  author = {Burba, D. and Mackoit-Sinkevi\ifmmode \check{c}\else \v{c}\fi{}ien\ifmmode \dot{e}\else \.{e}\fi{}, M. and Novi\ifmmode \check{c}\else \v{c}\fi{}enko, V. and Witkowska, E. and Juzeli\ifmmode \bar{u}\else \={u}\fi{}nas, G.},
  journal = {Phys. Rev. A},
  volume = {109},
  issue = {5},
  pages = {053319},
  numpages = {8},
  year = {2024},
  month = {May},
  publisher = {American Physical Society},
  doi = {10.1103/PhysRevA.109.053319},
  url = {https://link.aps.org/doi/10.1103/PhysRevA.109.053319}
}

@article{RevModPhys.94.041001,
  title = {Colloquium: Atomtronic circuits: From many-body physics to quantum technologies},
  author = {Amico, Luigi and Anderson, Dana and Boshier, Malcolm and Brantut, Jean-Philippe and Kwek, Leong-Chuan and Minguzzi, Anna and von Klitzing, Wolf},
  journal = {Rev. Mod. Phys.},
  volume = {94},
  issue = {4},
  pages = {041001},
  numpages = {30},
  year = {2022},
  month = {Nov},
  publisher = {American Physical Society},
  doi = {10.1103/RevModPhys.94.041001},
  url = {https://link.aps.org/doi/10.1103/RevModPhys.94.041001}
}

@article{shahi2016new,
  title={A new polarized neutron interferometry facility at the NCNR},
  author={Shahi, CB and Arif, M and Cory, DG and Mineeva, T and Nsofini, J and Sarenac, D and Williams, CJ and Huber, MG and Pushin, DA},
  journal={Nuclear Instruments and Methods in Physics Research Section A: Accelerators, Spectrometers, Detectors and Associated Equipment},
  volume={813},
  pages={111--122},
  year={2016},
  publisher={Elsevier}
}

@article{pushin2015neutron,
  title={Neutron interferometry at the national institute of standards and technology},
  author={Pushin, Dimitry A and Huber, Michael G and Arif, Muhammad and Shahi, Chandra B and Nsofini, Joachim and Wood, Christopher J and Sarenac, Dusan and Cory, David G},
  journal={Advances in High Energy Physics},
  volume={2015},
  number={1},
  pages={687480},
  year={2015},
  publisher={Wiley Online Library}
}

@article{Zehnder1891,
title={Ein neuer Interferenzrefraktor},
author={ Zehnder, L},
journal={Zeitschrift für Instrumentenkunde},
volume={11},
pages={275},
  year={1891}
}

@article{Mach1892,
title={Ueber einen Interferenzrefraktor},
author={Mach, L},
journal={Zeitschrift für Instrumentenkunde},
volume={12},
pages={89},
  year={1892}
}

@article{schoen2003precision,
  title={Precision neutron interferometric measurements and updated evaluations of the n- p and n- d coherent neutron scattering lengths},
  author={Schoen, K and Jacobson, David L and Arif, M and Huffman, PR and Black, TC and Snow, WM and Lamoreaux, Steve K and Kaiser, H and Werner, SA},
  journal={Physical Review C},
  volume={67},
  number={4},
  pages={044005},
  year={2003},
  publisher={APS}
}

@article{saggu2016decoupling,
  title={Decoupling of a neutron interferometer from temperature gradients},
  author={Saggu, Parminder and Mineeva, Taisiya and Arif, Muhammad and Cory, David G and Haun, Robert and Heacock, Ben and Huber, Michael G and Li, Ke and Nsofini, Joachim and Sarenac, Dusan and others},
  journal={Review of Scientific Instruments},
  volume={87},
  number={12},
  year={2016},
  publisher={AIP Publishing}
}

@article{morello2010single,
  title={Single-shot readout of an electron spin in silicon},
  author={Morello, Andrea and Pla, Jarryd J and Zwanenburg, Floris A and Chan, Kok W and Tan, Kuan Y and Huebl, Hans and M{\"o}tt{\"o}nen, Mikko and Nugroho, Christopher D and Yang, Changyi and Van Donkelaar, Jessica A and others},
  journal={Nature},
  volume={467},
  number={7316},
  pages={687--691},
  year={2010},
  publisher={Nature Publishing Group UK London}
}

@article{awschalom2013quantum,
  title={Quantum spintronics: engineering and manipulating atom-like spins in semiconductors},
  author={Awschalom, David D and Bassett, Lee C and Dzurak, Andrew S and Hu, Evelyn L and Petta, Jason R},
  journal={Science},
  volume={339},
  number={6124},
  pages={1174--1179},
  year={2013},
  publisher={American Association for the Advancement of Science}
}

@article{elzerman2004single,
  title={Single-shot read-out of an individual electron spin in a quantum dot},
  author={Elzerman, JM and Hanson, R and Willems van Beveren, LH and Witkamp, B and Vandersypen, LMK and Kouwenhoven, Leo P},
  journal={nature},
  volume={430},
  number={6998},
  pages={431--435},
  year={2004},
  publisher={Nature Publishing Group UK London}
}

@article{sponar2015weak,
  title={Weak values obtained in matter-wave interferometry},
  author={Sponar, Stephan and Denkmayr, Tobias and Geppert, Hermann and Lemmel, Hartmut and Matzkin, Alexandre and Tollaksen, Jeff and Hasegawa, Yuji},
  journal={Physical Review A},
  volume={92},
  number={6},
  pages={062121},
  year={2015},
  publisher={APS}
}

@article{denkmayr2017experimental,
  title={Experimental demonstration of direct path state characterization by strongly measuring weak values in a matter-wave interferometer},
  author={Denkmayr, Tobias and Geppert, Hermann and Lemmel, Hartmut and Waegell, Mordecai and Dressel, Justin and Hasegawa, Yuji and Sponar, Stephan},
  journal={Physical Review Letters},
  volume={118},
  number={1},
  pages={010402},
  year={2017},
  publisher={APS}
}

@article{lemmel2022quantifying,
  title={Quantifying the presence of a neutron in the paths of an interferometer},
  author={Lemmel, Hartmut and Geerits, Niels and Danner, Armin and Hofmann, Holger F and Sponar, Stephan},
  journal={Physical Review Research},
  volume={4},
  number={2},
  pages={023075},
  year={2022},
  publisher={APS}
}

@article{powerFET,
    author = {C.-C.~Tu and C.-L.~Hung and K.-B.~Hong and S.~Elangovan and W.-C.~Yu and Y.-S.~Hsiao and W.-C.~Lin and R.~Kumar and Z.-H.~Huang and Y.-H.~Hong and Y.-K.~Hsiao and R.-H.~Horng and B.-Y.~Tsui and T.-L.~Wu and J.-H.~He and and H.-C.~Kuo},
    title = {Industry perspective on power electronics for electric vehicles},
    journal = {Nat. Rev. Electr. Eng.} ,
    year = {2024},
     volume={1},
  pages={435}}

@article{wood2014quantum,
  title={Quantum correlations in a noisy neutron interferometer},
  author={Wood, Christopher J and Abutaleb, Mohamed O and Huber, Michael G and Arif, Muhammad and Cory, David G and Pushin, Dmitry A},
  journal={Physical Review A},
  volume={90},
  number={3},
  pages={032315},
  year={2014},
  publisher={APS}
}

\end{document}